\documentclass[12pt,preprint]{aastex}
\usepackage{amsmath}
\usepackage{color}
\shorttitle{MARVELS-5b: A Low-Mass Companions to HIP 67526}
\shortauthors{Jiang et al.}

\begin{document}
\title{Very Low Mass Stellar and Substellar Companions to Solar-like Stars From MARVELS IV: A Candidate Brown Dwarf or Low-Mass Stellar Companion to HIP 67526}
\author{Peng Jiang\altaffilmark{2,1,24}, Jian Ge\altaffilmark{1}, Phillip Cargile\altaffilmark{3}, Justin R. Crepp\altaffilmark{4},
Nathan De Lee\altaffilmark{1,3}, Gustavo F. Porto de Mello\altaffilmark{5,12}, Massimiliano Esposito\altaffilmark{6,7},
Let{\'i}cia D. Ferreira\altaffilmark{5,12}, Bruno Femenia\altaffilmark{6,7}, Scott W. Fleming\altaffilmark{8,9,1},
B. Scott Gaudi\altaffilmark{10}, Luan Ghezzi\altaffilmark{11,12}, Jonay I. Gonz{\'a}lez Hern{\'a}ndez\altaffilmark{6,7},
Leslie Hebb\altaffilmark{3}, Brian L. Lee\altaffilmark{1,13}, Bo Ma\altaffilmark{1}, Keivan G. Stassun\altaffilmark{3,14},
Ji Wang\altaffilmark{1}, John P. Wisniewski\altaffilmark{15}, Eric Agol\altaffilmark{13}, Dmitry Bizyaev\altaffilmark{16},
Howard Brewington\altaffilmark{16}, Liang Chang\altaffilmark{1}, Luiz Nicolaci da Costa\altaffilmark{11,12},
Jason D. Eastman\altaffilmark{10,17,18}, Garrett Ebelke\altaffilmark{16}, Bruce Gary\altaffilmark{3}, Stephen R. Kane\altaffilmark{19},
Rui Li\altaffilmark{1}, Jian Liu\altaffilmark{1}, Suvrath Mahadevan\altaffilmark{1,8,9}, Marcio A. G. Maia\altaffilmark{11,12},
Viktor Malanushenko\altaffilmark{16}, Elena Malanushenko\altaffilmark{16}, Demitri Muna\altaffilmark{20}, Duy Cuong Nguyen\altaffilmark{1},
Ricardo L. C. Ogando\altaffilmark{11,12}, Audrey Oravetz\altaffilmark{16}, Daniel Oravetz\altaffilmark{16}, Kaike Pan\altaffilmark{16},
Joshua Pepper\altaffilmark{3}, Martin Paegert\altaffilmark{3}, Carlos Allende Prieto\altaffilmark{6,7}, Rafael Rebolo\altaffilmark{6,21},
Basilio X. Santiago\altaffilmark{12,22}, Donald P. Schneider\altaffilmark{8,9}, Alaina C. Shelden Bradley\altaffilmark{16},
Thirupathi Sivarani\altaffilmark{1,23}, Stephanie Snedden\altaffilmark{16}, J. C. van Eyken\altaffilmark{19}, Xiaoke Wan\altaffilmark{1},
Benjamin A. Weaver\altaffilmark{20}, Bo Zhao\altaffilmark{1}}

\email{jpaty@mail.ustc.edu.cn}

\altaffiltext{1}{Astronomy Department, University of Florida, 211 Bryant Space Science Center, P. O. Box 112055, Gainesville, FL 32611, USA}
\altaffiltext{2}{Key Laboratory for Research in Galaxies and Cosmology, The University of Science and Technology of China, Chinese Academy of Sciences, Hefei, Anhui, 230026, China}
\altaffiltext{3}{Department of Physics and Astronomy, Vanderbilt University, Nashville, TN 37235, USA}
\altaffiltext{4}{Department of Physics, University of Notre Dame, 225 Nieuwland Science Hall, Notre Dame, IN 46556, USA}
\altaffiltext{5}{Universidade Federal do Rio de Janeiro, Observat{\'o}rio do Valongo, Ladeira do Pedro Ant{\^o}nio, 43, CEP: 20080-090, Rio de Janeiro, RJ, Brazil}
\altaffiltext{6}{Instituto de Astrof{\'i}sica de Canarias, C/V{\'i}a L{\'a}ctea S/N, E-38200 La Laguna, Spain}
\altaffiltext{7}{Departamento de Astrof{\'i}sica, Universidad de La Laguna, E-38205 La Laguna, Tenerife, Spain}
\altaffiltext{8}{Department of Astronomy and Astrophysics, The Pennsylvania State university, 525 Davey Laboratory, University Park, PA 16802, USA}
\altaffiltext{9}{Center for Exoplanets and Habitable Worlds, The Pennsylvania State University, University Park, PA 16802, USA}
\altaffiltext{10}{Department of Astronomy, The Ohio State University, 140 West 18th Avenue, Columbus, OH 43210, USA}
\altaffiltext{11}{Observat{\'o}rio Nacional, Rua General Jos{\'e} Cristino, 77, 20921-400 S{\~a}o Crist{\'o}v{\~a}o, Rio de Janeiro, RJ, Brazil}
\altaffiltext{12}{Laborat{\'o}rio Interinstitucional de e-Astronomia (LIneA), Rio de Janeiro, RJ 20921-400, Brazil}
\altaffiltext{13}{Department of Astronomy, University of Washington, Box 351580, Seattle, WA 98195-1580, USA}
\altaffiltext{14}{Department of Physics, Fisk University, 1000 17th Ave. N., Nashville, TN 37208, USA}
\altaffiltext{15}{Homer L Dodge Department of Physics \& Astronomy, University of Oklahoma, 440 W Brooks St, Norman, OK 73019, USA}
\altaffiltext{16}{Apache Point Observatory, P.O. Box 59, Sunspot, NM 88349-0059, USA}
\altaffiltext{17}{Las Cumbres Observatory Global Telescope Network, 6740 Cortona Drive, Suite 102, Santa Barbara, CA 93117, USA}
\altaffiltext{18}{Department of Physics Broida Hall, University of California, Santa Barbara, CA 93106, USA}
\altaffiltext{19}{NASA Exoplanet Science Institute, Caltech, MS 100-22, 770 South Wilson Avenue, Pasadena, CA 91125, USA}
\altaffiltext{20}{Center for Cosmology and Particle Physics, New York University, New York, NY, USA}
\altaffiltext{21}{Consejo Superior de Investigaciones Cient{\'i}ficas, Spain}
\altaffiltext{22}{Instituto de F{\'i}sica, UFRGS, Caixa Postal 15051, Porto Alegre, RS 91501-970, Brazil}
\altaffiltext{23}{Indian Institute of Astrophysics, II Block, Koramangala, Bangalore 560 034, India}
\altaffiltext{24}{LAMOST Fellow}

\begin{abstract}
We report the discovery of a candidate brown dwarf or a very low mass stellar companion (MARVELS-5b)
to the star HIP 67526 from the Multi-object APO Radial Velocity Exoplanet Large-area Survey (MARVELS).
The radial velocity curve for this object contains 31 epochs spread over 2.5 years.
Our Keplerian fit using a Markov Chain Monte Carlo approach,
reveals that the companion has an orbital period of $90.2695^{+0.0188}_{-0.0187}$ days, an eccentricity of $0.4375 \pm 0.0040$
and a semi-amplitude of $2948.14^{+16.65}_{-16.55}$ m s$^{-1}$. Using additional high-resolution spectroscopy, we find
the host star has an effective temperature $T_{\rm{eff}}=6004 \pm 34$ K, a surface gravity
$\log g$ [cgs] $=4.55 \pm 0.17$ and a metallicity [Fe/H] $=+0.04 \pm 0.06$.
The stellar mass and radius determined through the empirical relationship of Torres et al. (2010), yields
1.10$\pm$0.09 $M_{\sun}$ and 0.92$\pm$0.19 $R_{\sun}$. The minimum mass of MARVELS-5b is $65.0 \pm 2.9 M_{Jup}$,
indicating that it is likely to be either a brown dwarf or a very low mass star, thus occupying a
relatively sparsely-populated region of the mass function of companions to solar-type stars.
The distance to this system is 101$\pm$10 pc from the astrometric measurements of Hipparcos.
No stellar tertiary is detected in the high-contrast images taken by either FastCam
lucky imaging or Keck adaptive optics imaging, ruling out any star with mass greater than 0.2$M_{\sun}$
at a separation larger than 40 AU.

\end{abstract}


\section{Introduction}
Brown dwarfs (BDs; Basri 2000) are the star-like objects, which are not massive enough to
sustain stable hydrogen burning, but are sufficiently massive to fuse deuterium
(Chabrier et al. 2000; Spiegel et al. 2011). As a result, their luminosity and
temperature drop throughout their lifetimes (e.g. Burrows et al. 1997; Baraffe et al. 2003).
To date, over 800 BDs have been directly and indirectly discovered through a variety of methods
(e.g., Rebolo et al. 1995; Oppenheimer et al. 1995; Ruiz et al. 1997; Tinney et al. 1997;
Kirkpatrick et al. 1999, 2000, 2011; Marcy \& Bulter 2000; Mayor \& Udry 2000; Sahlmann et al. 2011).
Most of the known BDs are free-floating objects detected in the imaging surveys. These surveys seem to imply
a continuous distribution of masses through the hydrogen burning limit, with the abundance of BD rivaling that of
stars.

The radial velocity (RV) technique has been rapidly developed in last three decades, and has led to the first
discoveries of extrasolar planets around solar-like stars (Latham et al. 1989; Mayor \& Queloz 1995; Marcy \& Butler 1996).
Since the reflex RV semi-amplitudes induced by BD companions could be many hundreds of meters per second, which are
considerably larger than the signals induced by planetary companions, RV surveys should easily discover BD companions.
However, only 60 BD companions to solar-like stars in relatively short ($P < 10^4$ days) orbits have been identified
in all the previous RV surveys (e.g. Marcy \& Butler 2000; Mayor \& Udry 2000; Vogt et al. 2002; Sahlmann et al. 2011;
D\'{i}az et al. 2012). The distribution of masses for spectroscopic companions to solar-like stars
shows a clear deficit in the BD mass range (the ``brown dwarf desert''; Marcy \& Butler 2000),
quite in contrast to the surveys of free-floating BDs. Moreover, the statistical investigations
of stellar companion to solar-like stars have shown a paucity of companions with mass ratios
($q \equiv M_c/M_\star$) $<$ 0.2, suggesting that the short period BD desert extends
in mass toward the low-mass star regime (Pont et al. 2005; Burgasser et al. 2007; Bouchy et al. 2011;
Wisniewski et al. 2012).

The Multi-object APO Radial Velocity Exoplanet Large-area Survey (MARVELS), part of the Sloan Digital Sky
Survey III (SDSS-III; Eisenstein et al. 2011) program\footnote{http://www.sdss3.org/surveys/marvels.php},
monitors several thousands of stars in the magnitude range $V$=8--12 by visiting each star $\sim$24 times
over an 18-month interval with moderate RV precision (Ge et al. 2008; Ge et al. 2009; Ge \& Eisenstein 2009).
Currently, more than ten very low mass stellar and substellar companion candidates have been identified. In order to
confirm the discoveries and characterize them further, the MARVELS survey team made extensive follow-up observations,
including high precision RV monitoring, high-resolution spectroscopy, time-series photometry and high-contrast imaging.

High precision RV follow-up observations are useful to refine the orbital solutions and to detect additional lower
mass companions in the candidate systems. We also used the multi-epoch high-resolution spectroscopy
to rule out potential false alarms due to spectral contamination at the moderate resolving power of MARVELS
spectrograph. MARVELS-1b was announced as the first detection of BD candidates from MARVELS
(Lee et al. 2011). Further analysis of precise radial velocities made with the Hobby-Eberly Telescope (HET)
High Resolution Spectrograph initially suggested an interior giant planet in a 3:1 period commensuribility with
MARVELS-1b. However, the apparent RV residuals to a one-companion fit were later proved to be due to spectral
contamination by a stellar companion. This was determined by the identification of strong line bisector variations
(Wright et al. 2013). MARVELS-1 is actually
a face-on double-lined spectroscopic binary, instead of a single star with a BD companion. In another MARVELS
candidate BD system (TYC 3010-1494-1), a highly eccentric, double-lined spectroscopic binary star system
masqueraded as the RV signal of a single star orbited by a very low mass companion (Mack et al. 2013). 

Excluding these two false positive detections, three out of four published MARVELS discoveries have a
possible tertiary companion detected at wide separations in their systems. MARVELS-2b is likely to be a low-mass
stellar companion with a short period orbit around the F star TYC 2930-00872-1 and a stellar tertiary is identified
by analyzing the long-term trend in the RV curve (Fleming et al. 2012). For MARVELS-3b, a faint candidate tertiary
companion is detected in the Keck adaptive optics image, separated by $\sim$1\arcsec~from its host star (TYC 4110-01037-1;
Wisniewski et al. 2012). Ma et al. (2013) detected a faint point source at a separation of
$\sim$0.6\arcsec~from the host star of MARVELS-4b (TYC 2087-00255-1) through high-contrast imaging.
Future proper motion observations are necessary to resolve whether the offset objects are physically associated
with the host stars. Nevertheless, these results have encouraged the MARVELS team to keep assessing the multiplicity
for every future discovery in the survey.

Currently, there are about 60 BD companions to solar-like stars reported in the literature. The distribution of masses
of the companions exhibits a local minimum (the most ``arid'' part of the desert) in the mass range of $\sim$30--50 $M_{Jup}$
(Sahlmann et al. 2011; Ma \& Ge 2013). The tentative bimodal distribution of mass may indicate that there are two formation
mechanisms of BD and low-mass stellar companions: the low-mass BDs form by core accretion in protoplanetary disks; while more
massive companions form by gravitational collapse (Grether \& Lineweaver 2006; Sahlmann et al. 2011; Ma \& Ge 2013).
Moreover, the properties of host stars might also have an important impact on the formation of BD companions. Bouchy
et al. (2011) reported that Super-Jupiters, BDs, and low-mass M dwarf companions (10--100 $M_{Jup}$) to G-type
($T_{\rm eff} \la 6200$K) stars were apparently less common than similar companions to hotter stars. Compared to the
metallicity of the planet hosts (Santos et al. 2001; Valenti \& Fischer 2005; Johnson et al. 2010), the hosts of BD
companions are not that metal rich in general (Ma \& Ge 2013). Apparently the statistics of physical parameters are important
for us to understand the formation and evolution of low-mass companions. Therefore, the MARVELS team has taken pains to
follow up MARVELS candidates in order to collect a uniformly characterized sample for a meta-analysis.

In this paper, we report a candidate BD or a low-mass stellar companion (MARVELS-5b) to HIP 67526 with a period of $\sim$90 days
from MARVELS. In Section 2.1, we describe the RV measurements and solve for the spectroscopic orbital elements using
Markov Chain Monte Carlo (MCMC) analysis. We analyze the photometric data from SuperWASP and the astrometric data from Hipparcos
in Sections 2.2 \& 2.3, respectively. In Section 3.1, we determine precise stellar parameters for the
primary star. Using the stellar mass derived in Section 3.2, we then estimate the mass of the companion in Section 3.3.
The evolutionary state of the host star is studied in Section 3.4. The high-contrast imaging is presented in
Section 3.5. Finally, we provide a discussion and a summary in Section 4.

\section{Observations and Results for the Low Mass Companion}
\subsection{Differential Radial Velocities}
\subsubsection{MARVELS and TNG/SARG Measurements}
HIP 67526 was selected as an RV survey target according to the MARVELS
preselection criterion (Lee et al. 2011). It has been monitored at 21 epochs using the MARVELS
instrument mounted on SDSS 2.5-m Telescope at Apache Point Observatory (Gunn et al. 2006) during the first
two-year cycle of the SDSS-III MARVELS planet search program (Ge et al. 2008).
The MARVELS instrument is a fiber-fed dispersed fixed-delay interferometer instrument
capable of observing 60 objects simultaneously, designed for a large-scale RV survey (Ge 2002; Ge et al. 2009).
The dispersed fixed-delay interferometer instrument principle is described in
several prior papers (Ge 2002; Ge et al. 2002; Erskine 2003; Ge et al. 2006;
van Eyken et al. 2010; Wang et al. 2011). The MARVELS interferometer delay
calibration is described by Wang et al. (2012a, b). The interferometer produces two
fringing spectra per object, covering a wavelength range of 5000--5700~{\AA}, with resolving
power of $R\sim$12,000. Two iodine absorption spectra of light from a tungsten lamp taken before and
after each science exposure are used to calibrate any instrument drift.
Data processing and the error estimation algorithm have been described in detail
by Lee et al. (2011) and Fleming et al. (2010), respectively.

HIP 67526 was identified as a star bearing an unseen companion by performing Lomb-Scargle
(L-S) periodogram analysis (e.g. Lomb 1976; Scargle 1982 ;Cumming 2004; Baluev 2008) on the
21 MARVELS RV points. There are two significant peaks on the L-S periodogram with periods at $\sim$88 days
and $\sim$46 days (Figure 1). The false alarm probability (hereafter FAP)
of the 88 day peak is 0.00367\%, and the FAP of the 46 day peak is 0.0201\%. We fit a Keplerian orbit
to the observed RV curve, forcing the period to be close to $\sim$88 days
and $\sim$46 days. The preliminary fitting results are illustrated in Figure 2. The
solution at an orbital period of 90.2 days provides a better fit to the MARVELS RV curve than the
solution at an orbital period of 45.6 days. The shorter orbital period peak in the periodogram is
probably an alias. The minimum mass (if $\sin i=$1) of the unseen companion from
the longer period solution is $\sim$65 $M_{Jup}$ (see Section 3.3 for details). The estimated minimum mass
is below the hydrogen burning limit and places MARVELS-5b within the sparsely-populated region of the mass
function of companions to solar-like stars.

We collected ten additional RV measurements with the SARG spectrograph (Gratton et al. 2001)
at the 3.58m Telescopio Nazionale Galileo (TNG) Telescope in late 2010 and 2011.
The spectrograph covers a wavelength range of 4620--7920~{\AA} with $R\sim$57,000. The simultaneous
iodine cell technique (Butler et al. 1996) was employed to calibrate the RV
measurements. The raw spectra were reduced by using the standard IRAF\footnote{http://iraf.noao.edu/}
Echelle reduction packages. The final extracted differential RVs from MARVELS and TNG/SARG are presented
in Table 1. The RV curve was sampled in total at 31 epochs using these two instruments over 2.5 years.

\subsubsection{Spectroscopic Orbital Elements}
We have performed a Bayesian analysis of the observed radial velocities using a model
consisting of the primary star and one low-mass companion on an eccentric Keplerian orbit based on the combined
differential RV observations of MARVELS and TNG/SARG.

We calculated a posterior sample using the MCMC technique as described in Ford (2006).
Each state in the Markov chain is described by the parameter set
$\vec{\theta}=\{P,K,e,\omega,M,\gamma_M,\gamma_T,\sigma_j\}$, where $P$ is
orbital period, $K$ is the velocity semi-amplitude, $e$ is the orbital
eccentricity, $\omega$ is the argument of periastron, $M$ is the
mean anomaly at the chosen epoch ($\tau$). The parameters $\gamma_M$ and $\gamma_T$ are constant
systemic velocity terms for the MARVELS and TNG/SARG instruments respectively, used to account for
the offsets between the observed differential RV data and the zero point of the Keplerian RV model.
The ``jitter'' parameter, $\sigma_j$, describes any excess noise (Wright 2005), including both
astrophysical sources of noise (e.g. stellar oscillation, stellar spots) and any instrumental
noise not accounted for in the quoted measurement uncertainties.
We use standard priors for each parameter (see Ford \& Gregory 2007). The prior is uniform in the log
of the orbital period $P$,  while for $K$ and $\sigma_j$ we used a modified Jefferys prior (Gregory 2005).
Priors for the remaining parameters are uniform: $e$ (between zero and unity), $\omega$ and
$M$ (between zero and $2\pi$), $\gamma_M$ and $\gamma_T$. Following Ford (2006), we adopt a likelihood (i.e.,
conditional probability of making the specified measurements given a particular
set of model parameters) of
\begin{equation}
p(v|\vec{\theta},M) \propto \prod_k \frac{\exp[-(v_{k,\theta}-v_k)^2/2{\sigma_k}^2]}{\sqrt{{\sigma_j}^2+{\sigma_k}^2}},
\end{equation}
where $v_k$ is observed velocity at time $t_k$, $v_{k,\theta}$ is the
model velocity at time $t_k$ given the model parameters
$\vec{\theta}$, and $\sigma_k$ is the measurement uncertainty
for the observation at time $t_k$.

To test the robustness of the MCMC analysis, we calculate five Markov
chains starting from different initial states, each for
$5\times10^7$ states. To prevent the choice of initial states from
influencing our results, we consider only the second half of each
chain. We calculate the Gelman-Rubin test statistic (that compares the
variance of a parameter within each chain to the variance between chains; Gelman \& Rubin 1992)
for each model parameter. We find no indications that the Markov chains have yet to converge and conclude
that the Markov chains provide an adequate posterior sample for
inferring the orbital parameters and uncertainties.

We combine the Markov chains described above to estimate the joint
posterior probability distribution for the orbital model of
HIP 67526. For orbital eccentricity, we also used the $\Gamma$ method described in Wang (2011),
which leads to a similar result to that from the MCMC analysis.
The median values are taken for each model parameter based on the marginal
posterior probability distributions. The uncertainties
are calculated as the standard deviation about the mean value from
the combined posterior sample. Since the shape of the marginal
posterior distribution is roughly similar to a multivariate normal
distribution, the median value plus or minus the reported uncertainty
roughly corresponds to a 68.3\% confidence interval. Finally, we convert
the model parameters to traditional standard parameters of a spectroscopic
orbit and report the results in Table 2. The phase-folded RV curve is presented
in Figure 3.

\subsection{SuperWASP Photometry}
We searched the SuperWASP public archived database (Butters 2010) and found 1378 photometric data measurements of
HIP 67526 observed in 2004 and 5680 data points in 2007. The mean absolute deviation of the light curve is 9.7 mmag.
We first searched for a transit-like dip in brightness at short periods between 0.2--10 days. We find no
significant detection of a transit event. Next, we searched for transits specifically in the range of 85--95 days,
which includes the best-fit period from the spectroscopic RV curve. The phase-folded data are sparsely
covered at these long periods, and we find no significant transit signal.
In summary, we do not find a transit in SuperWASP photometric data with a long or a short period.
We also attempted to search for a sinusoidal signal in the light curve but found no significant signal.

\subsection{Hipparcos Astrometry}
HIP 67526 exists in the Hipparcos catalog with a parallax distance of 100$\pm$10 pc from the Sun.
It is possible that the orbital motion of the star due to the gravitational influence of its companion
can be resolved by Hipparcos astrometry. This would allow the the inclination $i$ and the ascending node $\Omega$
of the Keplerian orbit, and thus the true mass of MARVELS-5b, to be well constrained (Sahlmann et al. 2011).
We retrieved the dataset of HIP 67526
from the Intermediate Astrometric Data (IAD) of the new Hipparcos reduction (van Leeuwen 2007), including
the satellite orbit number, the epoch $t$, the parallax factor $\Pi$, the scan angle orientation $\psi$,
the abscissa residual $\delta \Lambda$ and the abscissa error $\sigma_{\Lambda}$ for every satellite scan.
There are 123 available Hipparcos scans on HIP 67526 in the IAD and the average abscissa error is
$\bar{\sigma_{\Lambda}} \sim$10 mas. Thus, the dataset allows a 1$\sigma$ detection of an orbit with an angular
size of $\bar{\sigma_{\Lambda}}/\sqrt{N} = 10/\sqrt{123} \sim 1$ mas.

We then estimate the minimum angular semimajor axis (in mas) of the primary's orbit, which can be written as
\begin{equation}
a_a\sin i = 3.35729138\times10^{-5}KP\sqrt{1-e^2}\varpi,
\end{equation}
where $K$ (in m s$^{-1}$), $P$ (in yr) and $e$ the spectroscopic orbital elements, $\varpi$ (in mas) the parallax,
$i$ the unknown inclination (Pourbaix 2001). This equation yields a minimum angular semimajor axis $\sim$ 0.2 mas for HIP 67526.
Therefore, for nearly edge-on orbits, the angular size of the primary's orbit is well below the 1$\sigma$ detection threshold,
and thus the motion of HIP 67526 about the system's center-of-mass cannot be detected for such geometries. Assuming that the
Hipparcos data of HIP 67526 are consistent with no astrometric signal from the orbit about the center-of-mass of the system,
and that orbits of $\sim 1$, $\sim 2$, and $\sim 3$ mas would have been detected at $1$, $2$, and $3\sigma$, we can place an
upper limits on the companion mass of $\sim 0.33~M_\sun$ ($1\sigma$), and $\sim 0.80~M_\sun$ ($2\sigma$), and $1.49~M_\sun$
($3\sigma$).  As argued in Section 3.3, such massive companions are anyway a posteriori unlikely for flat or falling priors
on the companion mass distribution.  For priors that increase with increasing mass, companions of mass $\ga 0.5~M_\sun$ are not a
posteriori implausible, but would be ruled out based on the lack of evidence of a second set of spectral lines in the high-resolution
spectra, if the companion was luminous (i.e., not a remnant)

\section{Observations and Results for the Host Star}
\subsection{Spectroscopic Parameters and Spectral Energy Distribution Analysis}
In order to characterize the host star HIP 67526, two moderate-resolution spectra (R$\sim$31,500) were taken
with the ARC Echelle Spectrograph (ARCES; Wang et al. 2003) mounted on the Apache Point Observatory 3.5m
telescope on UT 2010 June 10. The spectra cover the full optical range from 3600 \AA\ to 1.0 $\mu$m.
The spectra were obtained using the default $1.^{\prime \prime}6$ $\times$ $3.^{\prime \prime}2$ slit and an exposure time of 1200 s.
The raw data were processed using standard IRAF techniques. The extracted 1D spectra were converted to vacuum
wavelengths and to the heliocentric frame. The data were normalized by fitting a series of polynomials to the continuum.

We utilized two individual pipelines to derive basic stellar parameters such as $T_{\rm{eff}}$, $\log g$
and [Fe/H] for the host star. Both pipelines are based on the requirements of excitation and ionization equilibria
of Fe~I and Fe~II. However, different versions of ATLAS9 plane-parallel model atmospheres (Kurucz 1993 and Castelli
\& Kurucz 2004) and different iteration algorithms are implemented. We refer the readers to Wisniewski et al. (2012)
for more details on the pipelines. The derived stellar parameters from these two pipelines are usually
consistent to within 1$\sigma$ of the associated errors. Thus, we simply adopted the weighted average values as the final
determined stellar parameters. We combined the internal errors from the two pipelines as
$1/\sigma^2 = 1/\sigma_1^2 + 1/\sigma_2^2$ for each parameter, and added in quadrature a systematic error of
18 K, 0.08 ,0.03 and 0.02 km s$^{-1}$ for $T_{\rm{eff}}$, $\log g$, [Fe/H] and $V_{\rm {mic}}$, respectively
(Wisniewski et al. 2012). The final results are summarized in Table 3.

We collected the optical and NIR absolute photometry of HIP 67526 from the Hipparcos,
2MASS and WISE catalogs (Table 3) to construct a spectral energy distribution (SED; see Figure 4)
and fit it with a NextGen model atmosphere (Hauschildt et al. 1999).
The resultant stellar parameters, $T_{\rm{eff}} = 5800\pm200$ K, $\log g$ [cgs] $= 4.0\pm1.0$ and
[Fe/H] = $0.0\pm0.5$, are in good agreement with the parameters derived from spectroscopy
within the errorbars. In addition, the SED fitting indicates that HIP 67526 suffers only slight
extinction ($A_{\rm{V}} = 0.035\pm0.035$).

\subsection{Stellar Mass and Radius}
We determine the stellar mass and radius using two methods. First, we use the empirical relationship of Torres et al. (2010) with
our values for $T_{\rm{eff}}$, $\log g$ and [Fe/H]. Uncertainties in the mass and radius are derived by adding in quadrature
the correlations of the best-fit coefficients from Torres et al. (2010) and the scatter in the relation as reported in their study.
The correlations between the stellar parameters $T_{\rm{eff}}$, $\log g$ and [Fe/H] are not measured and are therefore not
considered. We find a mass $M_{\star} = 1.10 \pm 0.09 ~ M_\sun$ and a radius $R_{\star} = 0.92 \pm 0.19 ~ R_\sun$.

The existence of a trigonometric parallax provides additional information to constrain the mass and radius of the primary star.
We incorporate this data by running a MCMC analysis that fully explores parameter space.  One million iterations in
the MCMC chain were run, stepping through $T_{\rm eff}$, $\log g$, [Fe/H], parallax ($\varpi$) and $A_{\rm{V}}$.  We use random starting
values to initiate the chain. For each iteration, we calculate a mass and radius following Torres et al. (2010) and the iteration's values
of $T_{\rm {eff}}$, $\log g$ and [Fe/H]. A stellar luminosity is calculated via the Stefan-Boltzmann law, then a bolometric correction to
the 2MASS $\rm{K_{s}}$ band is applied by interpolating the table of corrections as a function of $T_{\rm eff}$ for [M/H] = 0.0 and
$\log g = 4.5$ from Masana et al. (2006). The absolute $\rm{K_{s}}$ magnitude is calculated from the luminosity and bolometric correction,
after which the apparent magnitude is calculated from the absolute magnitude and the iteration's values of $\varpi$ and $A_{\rm{V}}$.

After each iteration, a $\chi^2$ statistic is calculated as the sum of the individual $\chi^2$ for $T_{\rm eff}$, $\log g$, [Fe/H],
$\varpi$ and $A_{\rm{V}}$, where the expected values for $T_{\rm eff}$, $\log g$ and [Fe/H] are the values determined spectroscopically,
the expected value for $\varpi$ comes from the Hipparcos catalog, and the expected value for $A_{\rm{V}}$ comes from the SED analysis.
The next iteration's trial parameters are selected using Gaussians centered on the current iteration's values with widths equal to
the 1$\sigma$ parameter uncertainties for $T_{\rm eff}$, [Fe/H] and $A_{\rm{V}}$, and $0.1 \sigma$ for $\log g$ and $\varpi$. These
widths were empirically determined such that the overall trial acceptance rate was ${\sim}24$\%, close to the optimal value for
multi-dimensional chains (Gelman et al. 2003).

The first 1\% of iterations are rejected as a burn-in period, while the remaining iterations are used to determine the best-fit final
parameters ($M_{\star}$, $R_{\star}$, $T_{\rm eff}$, $\log g$, [Fe/H], $\varpi$, $A_{\rm{V}}$). The 1$\sigma$ uncertainties are derived based on the
cumulative histogram of each parameter. For the stellar mass and radius uncertainties, the reported scatter in Torres et al. (2010) is also
added in quadrature. Each parameter agrees to within 1$\sigma$ of the spectroscopic/SED/catalog values, and are tabulated in Table 3.

\subsection{Mass of the Candidate Low-Mass Companion}
Using the spectroscopic orbital elements from the RV fit, we can derive the mass function of the companion,
\begin{equation}
M_f \equiv \frac{(M_c \sin i)^3}{(M_{\star}+M_c)^2} = \frac{K^3 (1-e^2)^{3/2}P}{2\pi G},
\end{equation}
which is independent of the mass of the primary and the inclination of orbit. For MARVELS-5b, we obtain,
\begin{equation}
M_f = (1.742 \pm 0.026) \times 10^{-4} M_{\sun},
\end{equation}
where the uncertainty is essentially dominated by the uncertainty in $K$ (see Table 2).
Assuming $\sin i=1$, we derive its minimum mass $M_{min} = 65.0 \pm 2.9 M_{Jup}$.
The uncertainty here is dominated by the uncertainty in the primary mass (see Table 3).
We also find the minimum mass ratio of the companion $q_{min}=0.0560 \pm 0.0015$.

The true mass of the companion depends on the inclination of its orbit, which is unknown.
We can estimate the posterior probability distribution of the true mass, assuming an isotropic
distribution of orbits and adopting a prior for the distribution of the companion mass ratios.
We therefore consider three reasonable priors on the companion mass ratio of the form:
$dN/dq \propto q^{\alpha}$, where $\alpha = -1, 0, +1$ (e.g., Grether \& Lineweaver 2006).
The estimation was realized by using a Monte Carlo, which has been described in detail in
Fleming et al. (2010) and Lee et al. (2011). All sources of uncertainty from the mass function
and the primary mass have been considered appropriately. We draw values of $\cos i$ from a uniform
distribution and weight the resulting distribution by $q^{\alpha+1}$ in order to account for the
mass ratio prior. For $\alpha>0$, the a posteriori distribution does not converge. However, we
can rule out mass ratios $q>~1$ for main-sequence companions by the lack of a second set of spectral
lines in the high-resolution spectra. We therefore enforce $q\le1$, thus implicitly assuming the
companion is not a stellar remnant. The resultant cumulative distributions of the true companion mass are
presented in Figure 5, and we summarize the median mass as well as the transit probability for each
of our priors in Table 4. For $\alpha<0$, MARVELS-5b is more likely to be a true BD; for
$\alpha=0$ or $\alpha=1$, it is more likely to be a low-mass stellar companion.

\subsection{Evolutionary State of the Host Star}
We estimate the evolutionary state of the host star HIP 67526 by comparing the
measured stellar parameters with a Yonsei-Yale stellar
evolutionary track (Demarque et al. 2004) for an analogous star
with $M_{\star} = $ 1.10 $M_{\sun}$ and $[Fe/H] =$ 0.04. The
result is displayed in Figure 6. The dashed curves represent the
same evolutionary track but for stellar masses $\pm$0.08
$M_{\sun}$, which is the 1$\sigma$ uncertainty in the stellar
mass from the Torres et al. (2010) relation. The shaded region
indicates the 1$\sigma$ deviations in the evolutionary track. The
blue dots are the location of the star at different ages.
The evolutionary data suggest a young star, since most of the area
of the 1$\sigma$ ellipsoid lies either below or very close
to the ZAMS. However, using the APO spectroscopic data, we
measure the flux in the line cores of the Ca~II H and K lines and
calculate the activity index $\log (R'_{HK})$, yielding $-$4.9.
This value points to an age of at least $\sim$3 Gyr (e.g.,
Figure 11 of Mamajek \& Hillenbrand 2008).
Both the HR diagram and the HK activity levels, however, are poor age
discriminants in this range of parameters. Taken together, the
evolutionary and activity data points to a star no younger than
2-3 Gyr, and probably no older than the Sun, a range compatible
with both criteria within the rather large errors. This range
corresponds to our best estimate of the age of HIP 67526.

\subsection{Direct Imaging Search for Visual Companions}
\subsubsection{FastCam Lucky Imaging}
Lucky imaging (LI, observations taken at very high cadence to achieve
nearly-diffraction-limited images from a subsample of the total) was performed
using FastCam (Oscoz et al. 2008) on the 1.5 m TCS telescope at Observatorio del
Teide in Spain. The primary goal of these observations was to search for
companions at large separations which could contaminate spectroscopic observations
of the target masquerading as a systematic trend in the RV data (Fleming et al. 2012). The LI frames
were acquired on 3 April 2011, 5 May 2011 and 8 May 2011 in the $I$ band
and spanning $\sim 21$\arcsec $\times 21$\arcsec~on sky. On 3 April 2011 a total of
100,000 short exposure images, each corresponding to 35 msec exposure time; on
5 May 2011 a total of 45000 short exposure images, each corresponding to 35
msec exposure time, and on 8 May 2011 a total of 45000 short exposure images,
each corresponding to 50 msec exposure were acquired. The data were processed
using a custom IDL software pipeline. After identifying corrupted frames due to
cosmic rays, electronic glitches, etc., the remaining frames are bias corrected
and flat fielded.

Lucky image selection is applied using a variety of selection thresholds (best
$X$\%) based on the brightest pixel (BP) method.  The selected BP must be below
a specified brightness threshold to avoid selecting cosmic rays or other
non-speckle features. As a further check, the BP must be consistent with the
expected energy distribution from a diffraction speckle under the assumption of
a diffraction-limited PSF. The BPs of each frame are then sorted from
brightest to faintest, and the best $X$\% are then shifted and added to generate
a final image. In Figure 7, we show for the data collected in
April 2011 and in May 2011 the results of the LI selection and shift-and-add for
different LI thresholds ranging from considering only the best 1\% of the frames
up to including 80\% of the data. Each panel covers $\sim 5.5$\arcsec $\times
5.5$\arcsec~centered on HIP 67526. Restricting the LI selection to
the top percentage (i.e. the 1\% LI image) improves the angular resolution with
respect to choosing a lower threshold (i.e. the 80\% LI image) but at the cost
of higher noise at large distances from the target.

We follow the same procedure as in Femen{\'{\i}}a et al. (2011) to compute the
3$\sigma$ detectability ($\Delta m$) curves on each of the images whose
$\sim 5.5$\arcsec $\times 5.5$\arcsec~region
around HIP 67526 has been depicted in Figure 7: at a given angular distance
$\rho$ from HIP 67526 we identify all possible sets of small boxes of a size
larger but comparable to the FWHM of the PSF (i.e. $5 \times 5$ pixel
boxes). Only regions of the image showing structures easily recognizable as
spikes due to diffraction of the telescope spider and/or artifacts on the
read-out of the detector are dismissed. For each of the valid boxes on the
arc at angular distance $\rho$ the standard deviation of the image pixels within
the $5\time 5$-pixel boxes is computed. The value assigned to the $3\sigma$
detectability curve at $\rho$ is 3 times the mean value from the standard
deviations of all the eligible boxes at $\rho$. This procedure on each of the LI
\% thresholding values (in steps of 1\%) produces a detectability curve, while the
envelope of all the family of curves for a given night yields the best possible
detectability curve to be extracted from the whole data set.
These ``best LI curves'' for each of the three nights are depicted in Figure 8,
where we can see the data collected are of similar quality with the data on
May 8th providing slightly better contrast values. No stellar tertiary to HIP 67526
is detected above the ``best LI curves''.

\subsubsection{Keck Adaptive Optics Imaging}
To further assess the multiplicity of HIP 67526, we acquired high angular resolution images of
the star on 24 June, 2012 UT using NIRC2 (instrument PI: Keith Matthew) with the Keck II adaptive optics
(AO) system (Wizinowich et al. 2000). AO observations probe the immediate vicinity of host stars,
and generate deep contrast compared to lucky imaging (e.g., Fleming et al. 2012; Ma et al. 2013).
Furthermore AO observations are sensitive to objects with red colors given the nominal
1-3 $\mu$m wavelength operating range.

Our observations consist of dithered frames taken with the K' ($\lambda_c = 2.12 \mu$m) filter.
We used the narrow camera setting to provide fine spatial sampling of the NIRC2 point-spread function.
The total on-source integration time was 190 seconds. Images were processed using standard techniques
to replace hot pixel values, flat field the detector array, subtract thermal background noise,
and align and coadd frames.

Figures 9 and 10 show the final reduced AO image and corresponding contrast curve. No candidate
companions were noticed in individual raw frames or the final reduced image. Our diffraction-limited
observations rule out the presence of companions with $\Delta m_K < 5$ mag for separations beyond
0.25" and $\Delta m_K < 8$ mag for separations beyond 1.0" ($10\sigma$). We employ the empirical
mass luminosity relationships in Delfosse et al. (2000) to derive the upper mass limit of the undetected
companions; this analysis results in an upper mass limit 0.2$M_{\sun}$ for separations larger than 40 AU and
0.1$M_{\sun}$ for separations larger than 100 AU.

\section{Discussion and Summary}
The frequency of BD companions to solar-like stars at close and intermediate separations
is less than 1\% (Marcy \& Butler 2000), which is much less than the frequency of planetary
companions ($>$10\%, e.g. Howard et al. 2010; Mayor et al. 2011) and the frequency of
spectroscopic stellar binaries detected in RV surveys ($\sim$14\% e.g. Halbwachs et al. 2003).
The frequency of BD companions was recently updated to be $<$0.6\% by Sahlmann et al. (2011)
on the basis of the CORALIE planet-search sample. This result is more accurate, since
the authors ruled out companions having true masses in the stellar regime using the
Hipparcos astrometric measurements to determine the orbital inclinations.
Constraining the mass distribution of companions can provide an important observational
clue to distinguish the formation and evolution mechanism of planetary, BD and stellar
companions. The current mass distribution suggests that low-mass BD companions less than $\sim$30 $M_{Jup}$
are likely to form in protoplanetary disks, while companions more massive than $\sim$45 $M_{Jup}$ forms
via fragmentation (Grether \& Lineweaver 2006; Sahlmann et al. 2011; Ma \& Ge 2013).
The BD and low-mass stellar companion discoveries from MARVELS will result in a more precise determination
of the mass limits of core accretion and gravitational collapse.
MARVELS-5b contributes to constraining the shape of the massive BD--low mass star boundary.

Spectroscopic binaries generally show moderately eccentric orbits (e.g., Duquennoy \& Mayor 1991;
Raghavan et al. 2010). Ribas \& Miralda-Escud\'{e} (2007) reported a tentative trend that low-mass
planets ($M\sin i <$ 4 $M_{Jup}$) generally have lower eccentricity than high-mass
planets ($M\sin i >$ 4 $M_{Jup}$), having a similar eccentricity distribution as binary stars
(Figure 3 of Ribas \& Miralda-Escud\'{e} 2007). D\'{i}az et al. (2012) reported that most of the
BD companions in their sample exhibit a considerable orbital eccentricity, supporting the
eccentricity-mass trend. MARVELS-5b has a high eccentricity ($\sim$0.44), which is around the peak
of the eccentricity distribution of the observed BD and low-mass stellar companions
(Sahlmann et al. 2011; D\'{i}az et al. 2012). MARVELS-5b probably is a member of the main
population of these massive companions to solar-like stars in view of its eccentricity. Our
previous MARVELS discoveries (MARVELS-2,3,4,6b) all have an eccentricity lower than $\sim$0.2
(De Lee et al. 2013).

A stellar tertiary is likely to affect the formation and evolution of the substellar
companion to the primary. Observationally, Zucker \& Mazeh (2002) point out that planets found
in binaries may have a negative period-mass correlation rather than the positive correlation
between the masses and periods of the planets orbiting single stars. By studying a larger sample
(19 planets in a double or multiple star system), Eggenberger et al. (2004, 2007) showed that short-period
($P <$ 40 days) planets found in multiple star systems may follow a different period-eccentricity
distribution than the short-period planets around isolated stars.
These observations seem to indicate that the presence of a stellar companion alters the migration
and mass growth rates of planets (Kley 2001). Similar influences have been also observed on
close spectroscopic binaries in triple systems. Shorter period binaries are more likely to be in multiple-star
systems, i.e. $\sim$80\% for $P <$ 7 days versus $\sim$40\% for $P >$ 7 days (Tokovinin et al. 2006).
This significant difference suggests that the periods of close binary systems with triples were
efficiently decreased by angular momentum exchange with companions.

With masses between planetary companions and stellar components in spectroscopic binaries,
the formation and migration of BD and low-mass stellar
companions can certainly be affected by the presence of a tertiary as well. However, this problem has not been studied in a
statistical way, since the current BD and low-mass stellar companion sample is fairly small and no systematic survey of
stellar tertiaries for these companions has been conducted. Using high-contrast imaging, the MARVELS survey goes to great 
lengths to investigate the statistics of its own discoveries of low-mass companions in the presence/absence of a stellar
tertiary. As mentioned in Section 1, most of the previous MARVELS discoveries have a stellar
tertiary (or a candidate stellar tertiary) detected by either the high-contrast imaging or the analysis of
long-term RV trend. Among the confirmed discoveries, MARVELS-3b (Wisniewski et al. 2012) has a similar orbital
period ($P \sim 79$ days) and minimum mass ratio ($q_{min} \sim 0.09$) as MARVELS-5b (this work), but the former has a
less eccentric orbit ($e\sim0.1$). Wisniewski et al. (2012) found a faint candidate tertiary companion on the Keck
adaptive optics image, separated by $\sim$1\arcsec~from the primary, thus speculating that MARVELS-3b might initially
have formed in a tertiary system with much different orbital parameters and reach its current short-period orbit during the cluster
dispersal phase. For MARVELS-5b, the Keck adaptive optics imaging rules out any star with mass greater than 0.1$M_{\sun}$
at a separation larger than 1\arcsec~from the primary. This may imply that other formation mechanisms of low-mass ratio binaries
are needed.

In summary, we report a candidate BD or low-mass stellar companion to the solar-like star HIP 67526.
The best Keplerian orbital fit parameters were found to have an orbital period of $90.2695^{+0.0188}_{-0.0187}$ days,
an eccentricity of $0.4375 \pm 0.0040$ and a semi-amplitude of $2948.14^{+16.65}_{-16.55}$ m s$^{-1}$. The minimum
companion mass was determined to be $65.0 \pm 2.9 M_{Jup}$. This object helps to populate the high-mass end of
the sparsely-populated region of the mass function of companions to solar-type stars and provide observational
evidence to constrain formation and evolution theories. No stellar tertiary is detected with high-contrast imaging
for the MARVELS-5 system, while all the other previous MARVELS discovered systems appear to have at least one
stellar companion.

\acknowledgments
Funding for the MARVELS multi-object Doppler instrument was provided by the W.M. Keck
Foundation and NSF with grant AST-0705139. The MARVELS survey was partially funded by
the SDSS-III consortium, NSF grant AST-0705139, NASA with grant NNX07AP14G and the
University of Florida. Funding for SDSS-III has been provided by the Alfred P. Sloan Foundation,
the Participating Institutions, the National Science Foundation, and the U.S. Department
of Energy. The SDSS-III web site is http://www.sdss3.org/. SDSS-III is managed by the
Astrophysical Research Consortium for the Participating Institutions of the SDSS-III Collaboration
including the University of Arizona, the Brazilian Participation Group, University of Cambridge,
University of Florida, the French Participation Group, the German Participation Group, the
Michigan State/Notre Dame/JINA Participation Group, Johns Hopkins University, Lawrence Berkeley
National Laboratory, Max Planck Institute for Astrophysics, New Mexico State University, New
York University, the Ohio State University, University of Portsmouth, Princeton University,
University of Tokyo, the University of Utah, Vanderbilt University, University of Virginia,
University of Washington and Yale University.

This work has made use of observations taken with the Telescopio Nationale Galileo (TNG)
operated on the island of La Palma by the Foundation Galileo Galilei, funded by the Instituto
Nazionale di Astrofisica (INAF), in the Spanish {\it Observatorio del Roque de los Muchachos}
of the Instituto de Astrof{\'i}sica de Canarias (IAC).

We have used data from the WASP public archive in this research. The WASP consortium comprises
of the University of Cambridge, Keele university, University of Leicester, The Open University,
The Queen's University Belfast, St. Andrews University and the Isaac Newton Group. Funding for
WASP comes from the consortium universities and from the UK's Science and Technology Facilities
Council. The publication makes use of data products from the Two Micron All Sky Survey, which
is a joint project of the University of Massachusetts and the Infrared Processing and Analysis
Center/California Institute of Technology, funded by the National Aeronautics and Space Administration
and the National Science Foundation.

P.J. acknowledges supports from Natural Science Foundation of China with grants NSFC 11233002,
NSFC 11203022 and the Fundamental Research Funds for the Central Universities. This research is
partially supported by funding from the Center for Exoplanets and Habitable Worlds. The Center for
Exoplanets and Habitable Worlds is supported by the Pennsylvania State University, the Eberly College
of Science, and the Pennsylvania Space Grant Consortium. Keivan Stassun, Leslie Hebb, and Joshua
Pepper acknowledge funding support from the Vanderbilt Initiative in Data-Intensive Astrophysics
(VIDA) from Vanderbilt University, and from NSF Career award AST-0349075. E.A. thanks NSF for
CAREER grant 0645416. G.F.P.M. acknowledges financial support from CNPq grant n$^o$ 476909/2006-6 and
FAPERJ grant n$^o$ APQ1/26/170.687/2004. L.G. acknowledges financial support provided by the PAPDRJ
CAPES/FAPERJ Fellowship. J.P.W. acknowledges support from NSF Astronomy \& Astrophysics Postdoctoral
Fellowship AST 08-02230. L.D.F. acknowledges financial support from CAPES. Work by B.S.G. was supported
by NSF CAREER Grant AST-1056524.

\begin{deluxetable}{lcrr}
\tabletypesize{\scriptsize}
\tablecaption{Differential Radial Velocity Measurements\label{tbl-1}}
\tablewidth{0pt}
\tablehead{
\colhead{Instrument} & \colhead{HJD$^a$} & \colhead{RV} & \colhead{Error} \\
\colhead{} & \colhead{ } & \colhead{(m s$^{-1}$)} & \colhead{(m s$^{-1}$)}
}
\startdata
MARVELS & 2454901.933194 & 2707.27 & 59.93 \\
        & 2454902.932454 & 2593.08 & 36.46 \\
        & 2454905.940208 & 2279.35 & 37.21 \\
        & 2454906.909213 & 2194.67 & 33.47 \\
        & 2454907.900972 & 2072.03 & 26.32 \\
        & 2454926.884479 & $-$1290.14 & 71.20 \\
        & 2454927.893461 & $-$1360.84 & 58.21 \\
        & 2454983.735220 & 3481.73 & 29.43 \\
        & 2454984.726944 & 3387.25 & 33.10 \\
        & 2455014.633588 & $-$864.37 & 33.17 \\
        & 2455023.635463 & $-$61.99 & 30.11 \\
        & 2455024.635984 & 414.89 & 38.37 \\
        & 2455222.898646 & 4441.05 & 36.96 \\
        & 2455223.889306 & 4515.50 & 33.08 \\
        & 2455254.894387 & 3393.48 & 33.77 \\
        & 2455258.906053 & 3081.05 & 31.48 \\
        & 2455259.892963 & 3006.84 & 29.68 \\
        & 2455260.882778 & 2900.49 & 33.60 \\
        & 2455261.939502 & 2788.23 & 31.63 \\
        & 2455280.845833 & 20.73 & 37.47 \\
        & 2455289.750139 & $-$1401.23 & 33.69 \\
\\
\cline{1-4}
TNG/SARG & 2455553.785137 & $-$3334.74 & 8.97 \\
         & 2455580.668228 & 1340.35 & 8.38 \\
         & 2455580.692489 & 1359.08 & 10.57 \\
         & 2455666.497671 & 873.73 & 9.94 \\
         & 2455666.521154 & 878.94 & 10.24 \\
         & 2455698.415610 & 998.61 & 9.27 \\
         & 2455725.369700 & $-$1587.03 & 10.94 \\
         & 2455760.380520 & 1250.09 & 10.09 \\
         & 2455791.354961 & 779.89 & 12.41 \\
         & 2455791.378282 & 760.40 & 13.30 \\
\enddata
\tablenotetext{a}{Heliocentric Julian Day}
\end{deluxetable}

\begin{deluxetable}{lcc}
\tabletypesize{\normalsize}
\tablecaption{Orbital Elements of MARVELS-5b\label{tbl-2}}
\tablewidth{0pt}
\tablehead{
\colhead{Parameter} & \colhead{Units} & \colhead{Value}
}
\startdata
$P$ & Period (days) & 90.2695$^{+0.0188}_{-0.0187}$ \\
$K$ & RV semi-amplitude (m s$^{-1}$) & 2948.14$^{+16.65}_{-16.55}$ \\
$e$ & Eccentricity & 0.4375$\pm$0.0040 \\
$\omega$ & Argument of periastron (degrees) & $-$140.91$\pm$0.54 \\
$T_0$ & Epoch of periastron (HJD) & 2455563.73$\pm$0.20 \\
$\gamma_M$ & MARVELS systemic velocity (m s$^{-1}$) & 2573.71$\pm$11.31 \\
$\gamma_T$ & TNG/SARG systemic velocity (m s$^{-1}$) & $-$374.69$\pm$11.4 \\
$\sigma_j$ & Jitter (m s$^{-1}$) & 21.35$^{+6.47}_{-4.97}$ \\
\enddata
\end{deluxetable}

\begin{deluxetable}{lcc}
\tabletypesize{\normalsize}
\tablecaption{Stellar Parameters of HIP 67526\label{tbl-3}}
\tablewidth{0pt}
\tablehead{
\colhead{Parameter} & \colhead{Result} & \colhead{Note}
}
\startdata
$B$ & 10.303$\pm$0.032 mag& Kharchenko \& Roeser (2009) \\
$V$ & 9.706$\pm$0.027 mag&  Kharchenko \& Roeser (2009) \\
$J$ & 8.598$\pm$0.020 mag& 2MASS \\
$H$ & 8.363$\pm$0.049 mag& 2MASS \\
$K$ & 8.295$\pm$0.024 mag& 2MASS \\
$W1$ & 8.226$\pm$0.022 mag& WISE \\
$W2$ & 8.283$\pm$0.020 mag& WISE \\
$W3$ & 8.260$\pm$0.017 mag& WISE \\
$W4$ & 8.063$\pm$0.150 mag& WISE \\
$T_{\rm eff}$ & 6004$\pm$34 K& Spectroscopy\\
$\log g$ [cgs] & 4.55$\pm$0.17 & Spectroscopy\\
{[Fe/H]} & $+0.04\pm0.06$ & Spectroscopy\\
$V_{\rm {mic}}$ & 1.03$\pm$0.04 km s$^{-1}$& Spectroscopy\\
$M_{\star}$ & 1.10$\pm$0.09 $M_{\sun}$& Torres et al. (2010) \\
$R_{\star}$ & 0.92$\pm$0.19 $R_{\sun}$& Torres et al. (2010)\\
$A_{\rm V}$ & 0.035$\pm$0.035 mag& SED Fitting\\
$\varpi$ & 9.87 $\pm$1.26 $\rm{mas}$ & Hipparcos\\
\\
\cline{1-3}
$M_{\star}$ & $1.11 \pm 0.08$ $M_{\sun}$ & MCMC\\
$R_{\star}$ & $0.95 ^{+0.15}_{-0.14}$ $R_{\sun}$ & MCMC\\
$T_{\rm eff}$ & $6004 \pm 34$ K& MCMC\\
$\log g$ [cgs] & $4.53 ^{+0.15}_{-0.13} $ & MCMC\\
{[Fe/H]} & $+0.04 \pm 0.06$ & MCMC\\
$\varpi$ & $10.25 ^{+1.09}_{-1.10}$ $\rm{mas}$ & MCMC\\
$A_{\rm{V}}$ & $0.043 ^{+0.033}_{-0.027}$ mag & MCMC\\
\enddata
\end{deluxetable}

\begin{deluxetable}{lrc}
\tabletypesize{\normalsize}
\tablecaption{The Companion Mass for Different Priors\label{tbl-4}}
\tablewidth{0pt}
\tablehead{
\colhead{Assumed Prior} & \colhead{Median Mass} & \colhead{Transit Prob.}
}
\startdata
None ($\sin i = 1$) & 65.0 $M_{Jup}$ & 100\% \\
$dN/dq \propto q^{-1}$ & 75.6 $M_{Jup}$ & 1.1\% \\
$dN/dq =$ const & 95.3 $M_{Jup}$ & 0.7\% \\
$dN/dq \propto q^{+1}$ & 243.8 $M_{Jup}$ & 0.2\% \\
\enddata
\end{deluxetable}
\clearpage

\begin{figure}
\epsscale{1}
\plotone{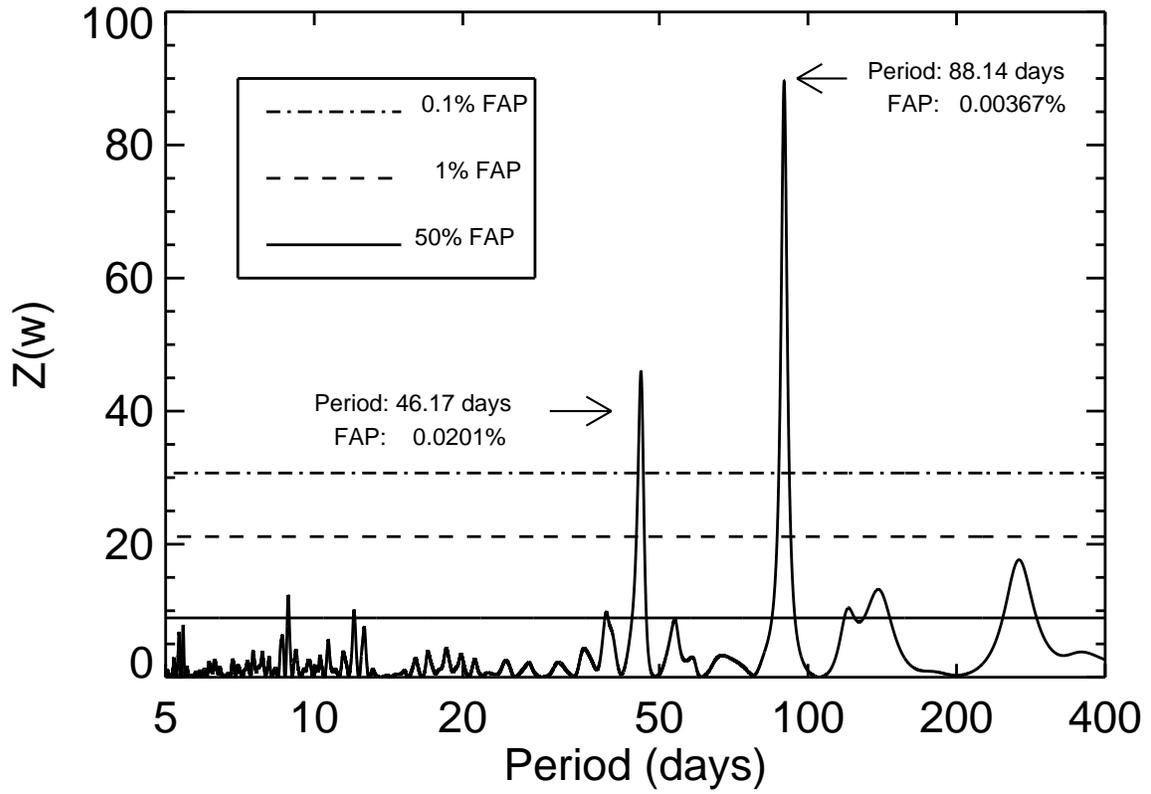}
\caption{\label{fig1} The periodogram for MARVELS RV measurements of HIP 67526
exhibits two peaks at periods of $\sim$88 days and $\sim$46 days. The three horizontal
lines indicate the false alarm probability at different levels (50\%, 1\%, 0.1\%).}
\end{figure}

\begin{figure}
\epsscale{1}
\plotone{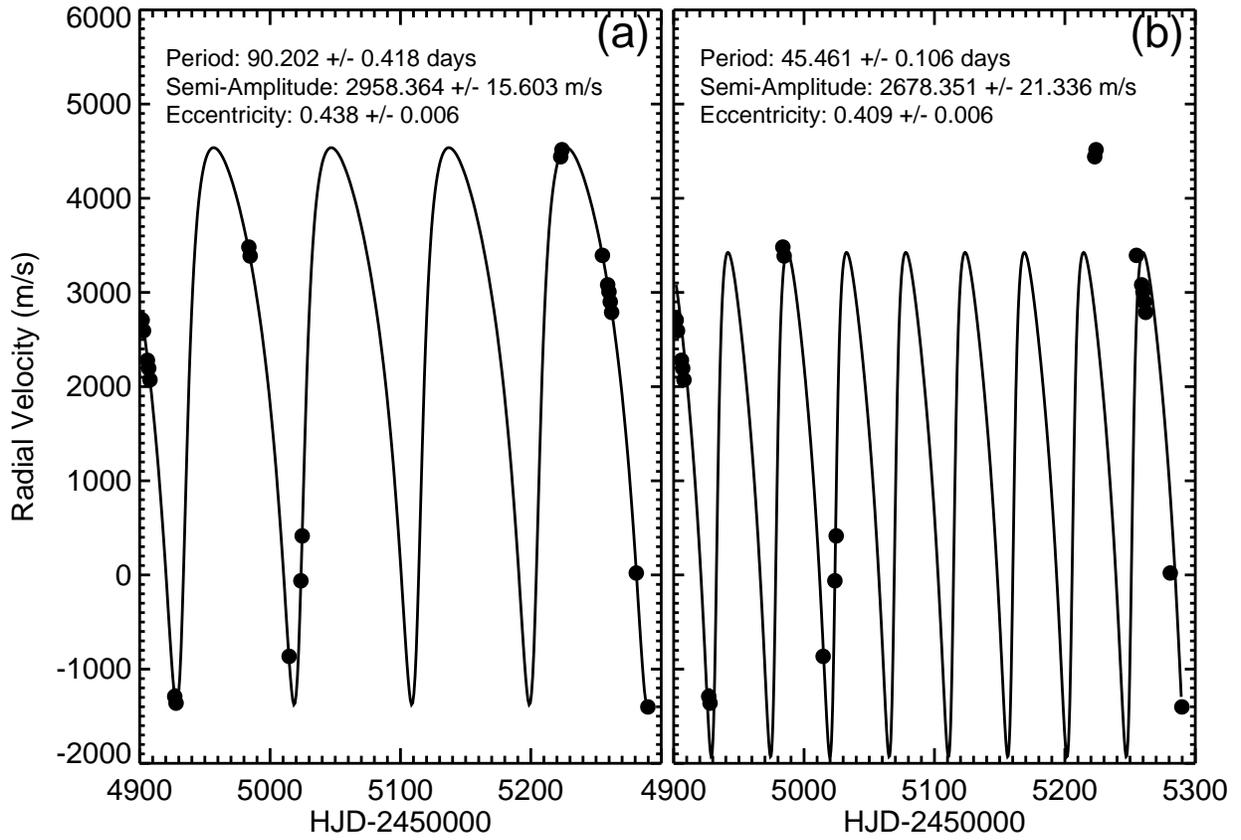}
\caption{\label{fig2} The Keplerian fitting results for the MARVELS RV measurements of HIP 67526
by forcing the period close to $\sim$88 days (panel a) and $\sim$46 days (panel b).}
\end{figure}

\begin{figure}
\epsscale{1}
\plotone{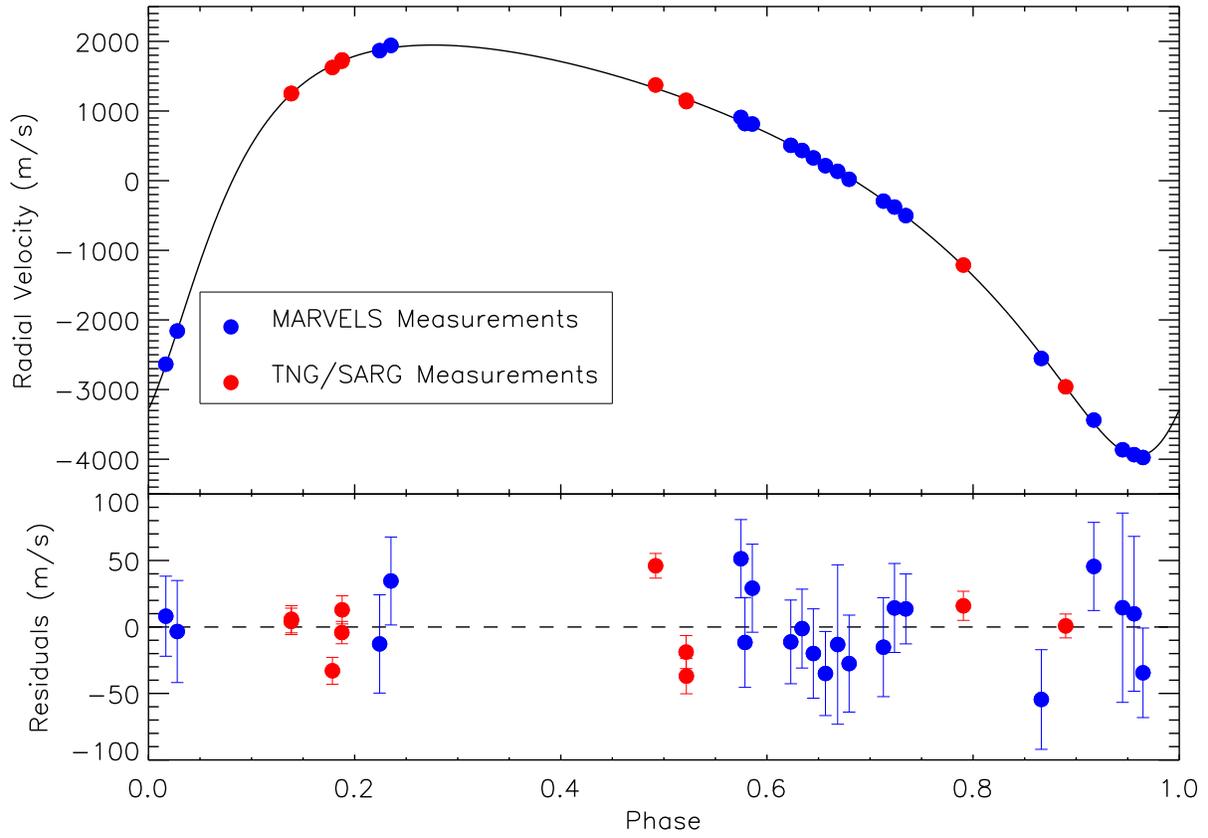}
\caption{\label{fig3} Phase-folded RV curve for MARVELS-5b, having
a period of $90.2695^{+0.0188}_{-0.0187}$ days, an eccentricity of $0.4375 \pm 0.0040$
and a semi-amplitude of $2948.14^{+16.65}_{-16.55}$ m s$^{-1}$. The blue dots are the
measurements from MARVELS and the red dots are the ones from TNG/SARG. Residuals
to this fit are shown in the bottom panel.}
\end{figure}

\begin{figure}
\begin{center}
\includegraphics[width=12cm, angle=90]{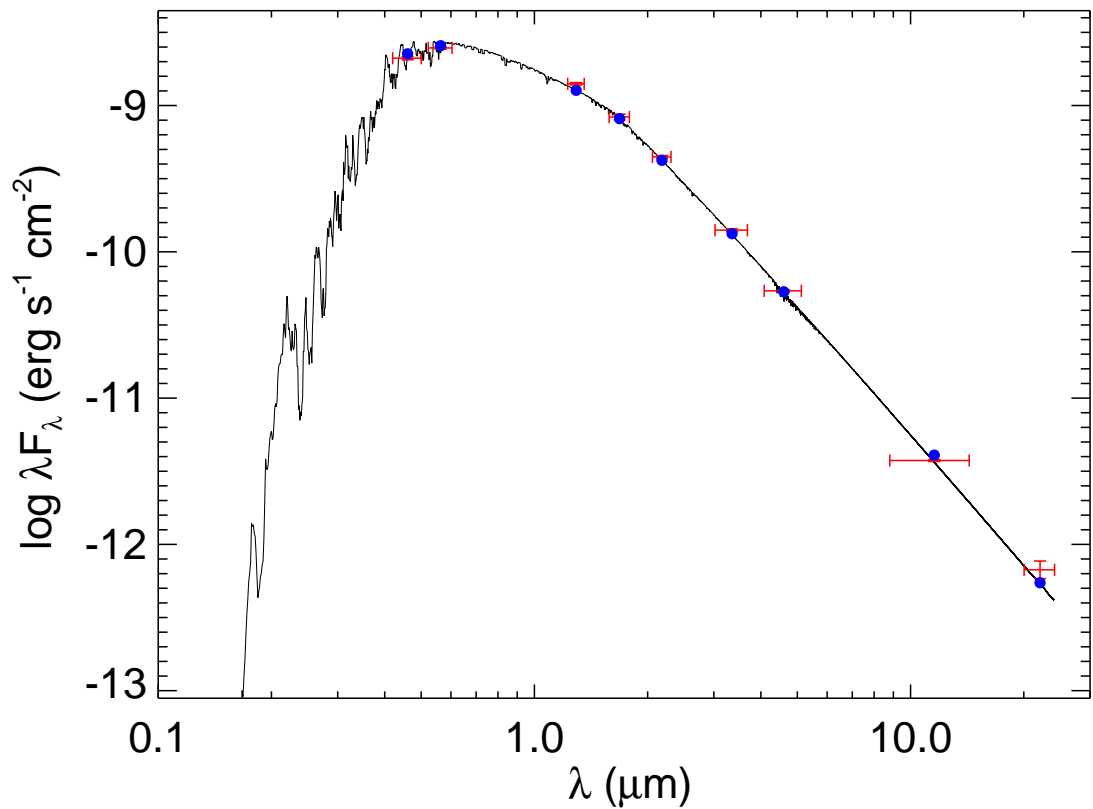}
\caption{The observed SED for HIP 67526 is overplotted with
the best-fit NextGen model atmosphere emission. Blue points
represent the expected fluxes in each band based on the best-fit
model, red horizontal bars are the bandpass widths, and red
vertical bars are the uncertainties of measured fluxes. The resultant
stellar parameters from this fit agreed to within 1 $\sigma$ with the
stellar parameters determined from analysis of moderate-resolution ARCES spectra.}
\label{fig4}
\end{center}
\end{figure}

\begin{figure}
\epsscale{1}
\plotone{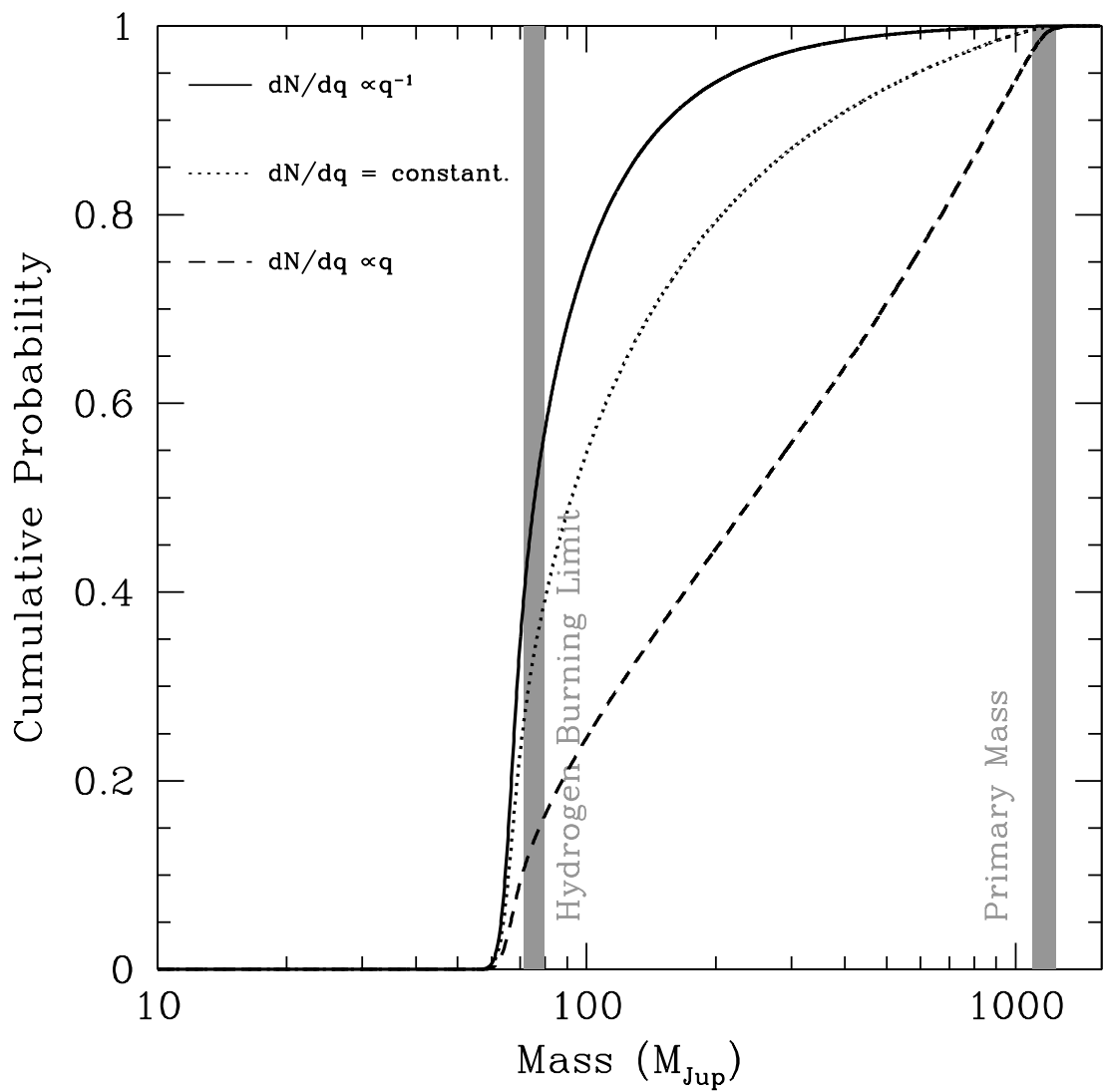}
\caption{\label{fig5}Cumulative probability that the mass of MARVELS-5b is less than
a given mass is shown, for three priors on the companion mass ratio: $dN/dq \propto q^{-1}$ (solid line),
$dN/dq =$constant (dotted line) and $dN/dq \propto q^{+1}$ (dashed line).}
\end{figure}

\begin{figure}
\begin{center}
\includegraphics[width=12cm, angle=90]{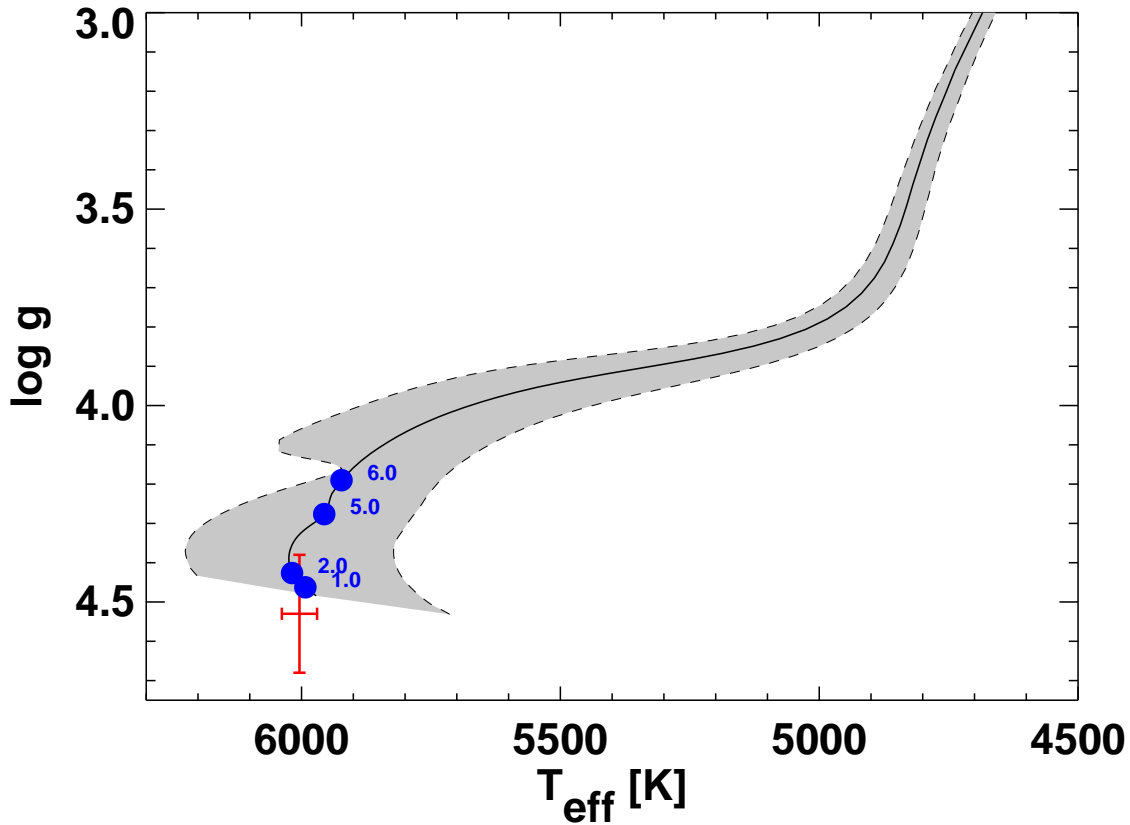}
\caption{The comparison of the observed stellar parameters of HIP
67526 with a Yonsei-Yale stellar evolutionary track (Demarque et
al. 2004) for an analogous star with $M_{\star} = $ 1.10
$M_{\sun}$ and $[Fe/H] = +0.04$. The shaded region indicates the
1$\sigma$ deviations in the evolutionary track. The blue dots are
the location of the analogous star at different ages in Gyr. HIP 67526
(in red) is most likely to be a main sequence dwarf star younger
than $\sim$ 2.5 Gyr, judging by the evolutionary data alone,
since most of the area within the 1$\sigma$ ellipsoid lies close
to the ZAMS. But its low level of activity suggests
an age over $\sim$ 3 Gyr, and thus it is most likely a
middle-aged star.} \label{fig6}
\end{center}
\end{figure}

\begin{figure}
\includegraphics[width=12cm]{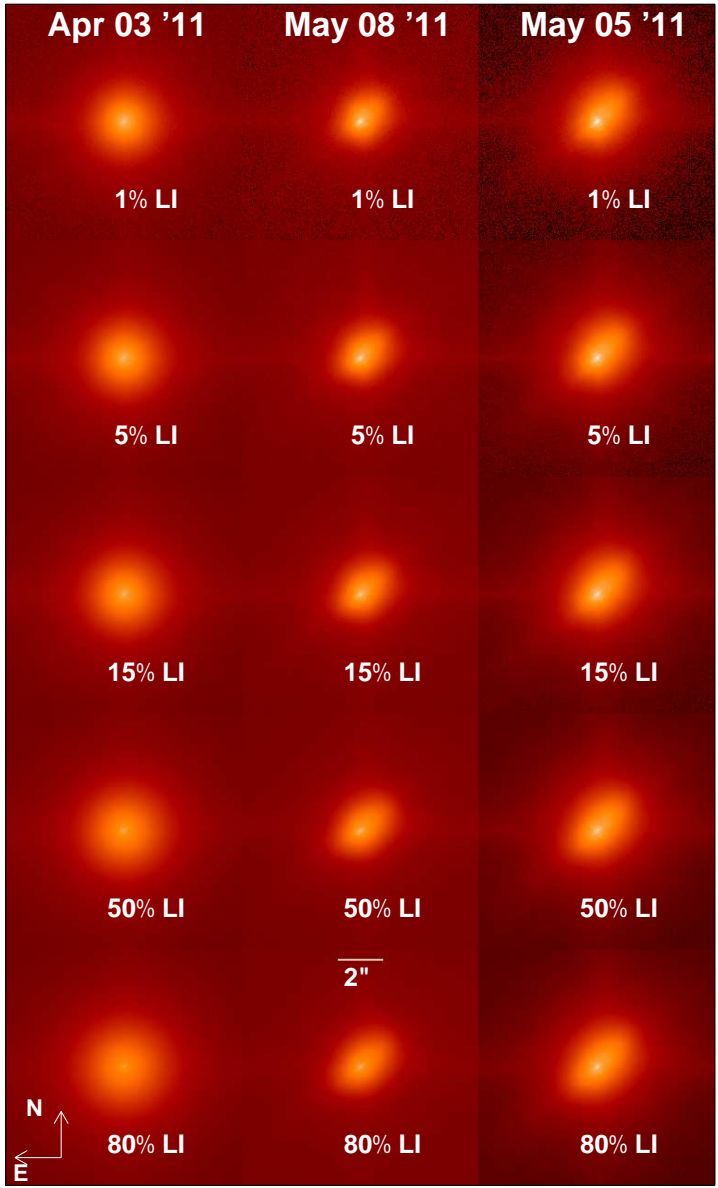}
\caption{Composite image showing the results of different LI thresholding on the
  frames acquired with FastCam at the TCS telescope on 3 April 2011, 5 May
  2011 and 8 May 2011. This set of images (in logarithmic scale) illustrates
  the gain in angular resolution close to the target location when applying high
  restrictive LI thresholds but at the cost of lowering the contrast achieved at
  large angular distances from target location (see also
  Fig.~\ref{fig:Detectability}).\label{fig:LI_panels}}
\end{figure}

\begin{figure}
\includegraphics[width=12cm, angle=90]{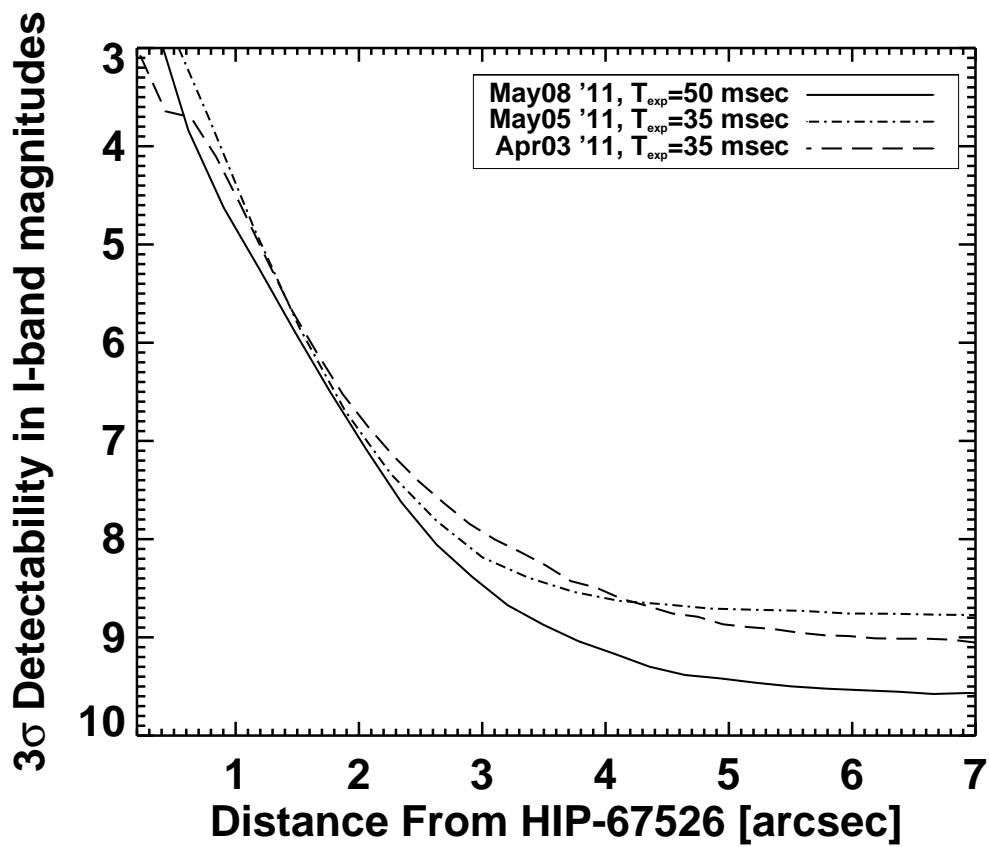}
\caption{Comparison of the best LI curves achieved on 3 April 2011, 5 May and 8 May 2011.
The 3$\sigma$ detectability ($\Delta m_I$) curves for individual nights were first computed on the images
obtained at different LI thresholds. The best LI curves are the envelope of all detectability
curves computed in steps of 1\% of LI thresholding}.
\label{fig:Detectability}
\end{figure}

\begin{figure}
\epsscale{1}
\plotone{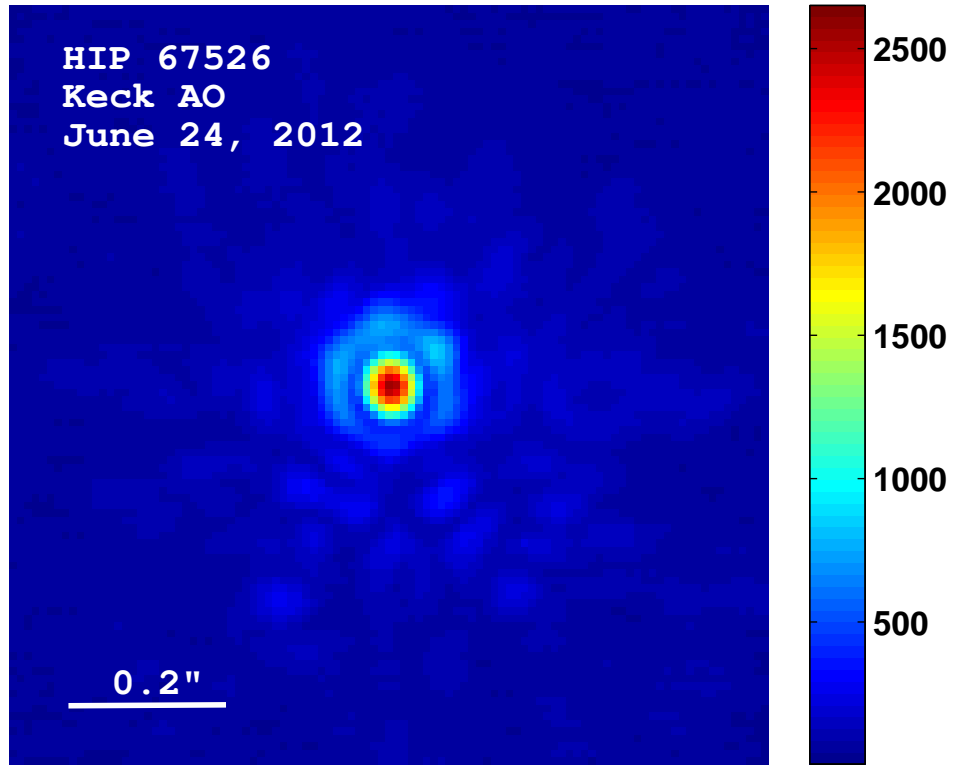}
\caption{\label{fig9} Keck AO image of HIP 67526. No stellar companions are detected
with $\Delta m_K < 5$ mag for separations beyond 0.25" and $\Delta m_K < 8$ mag for
separations beyond 1.0" at $10\sigma$ significance level (see also
  Fig.~\ref{fig10}).}
\end{figure}

\begin{figure}
\epsscale{1}
\plotone{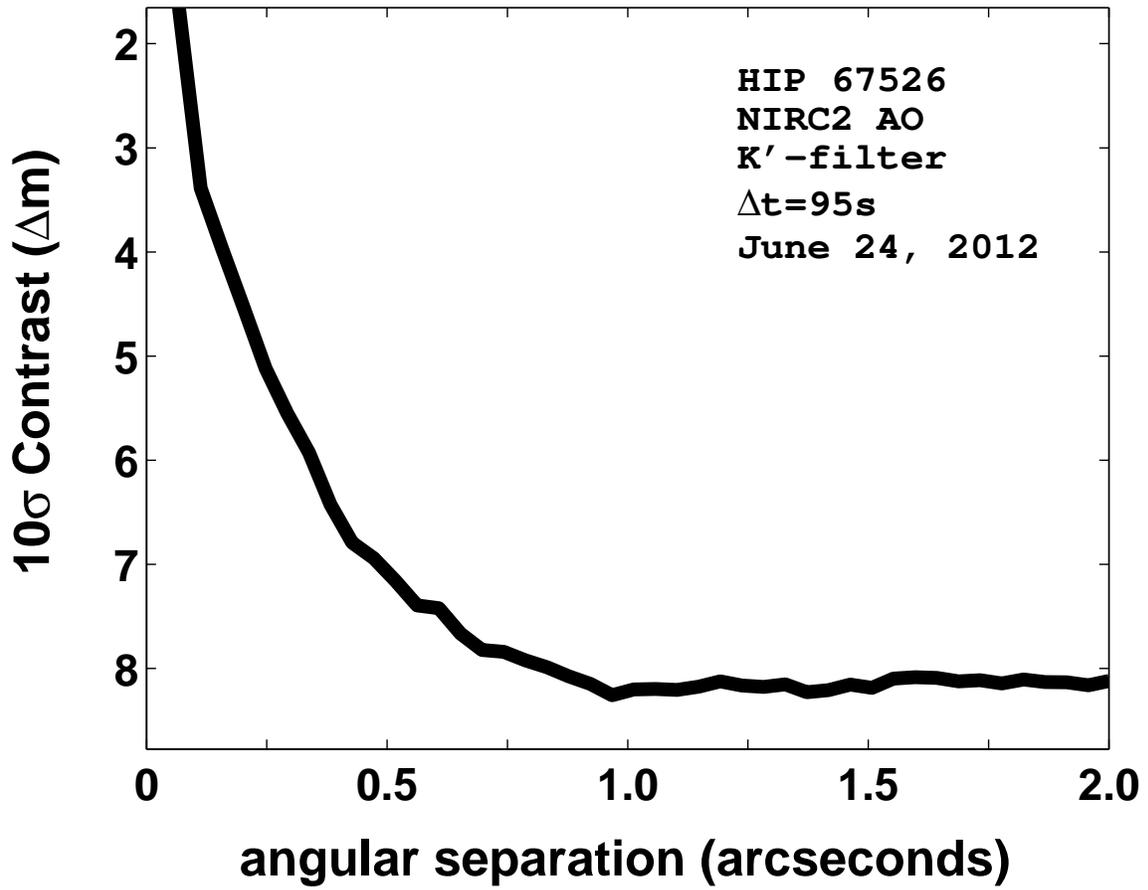}
\caption{\label{fig10} Detectability (contrast curve) for the Keck AO image of HIP 67526.
}
\end{figure}
\end{document}